# Gravitational Radiation and its Application to Space Travel, principles and required scientific developments.

Giorgio Fontana

*University of Trento, Department of Materials Engineering, 38050 Mesiano, TN, Italy*
*fontana@ing.unitn.it*

**Abstract.** Gravitational radiation is an elusive form of radiation predicted by general relativity, it is the subject of intense theoretical and experimental research at the limit of the sensitivity of today's instrumentation. In spite of the fact that no direct evidence of this radiation now exist, observed astrophysical phenomena have given convincing proofs of its existence. Theories predict that gravitational radiation may also be employed for propulsion, moreover the nonlinear behaviour of spacetime may permit the generation of spacetime singularities with colliding beams of gravitational radiation, this phenomenon could become a form of propellantless propulsion. Both applications would require gravitational wave generators with high power and appropriate optical properties. Among the proposed techniques that could be applicable to the production of gravitational waves, a promising one is the possible emission of gravitons by quantum systems. A hypothesis describing the production of gravitons in s-wave/d-wave superconductor junction is presented.

## INTRODUCTION

Everybody can have a personal experience with mass, gravity, inertia, and the effects of a reaction force. From this point of view, modern space propulsion is high technology applied to very old concepts.

Propulsion by reaction is a well established technology but it is not suitable for interstellar space travel because the total amount of propellant required would become unacceptable, moreover the related speed limitations would require missions lasting many decades if not centuries.

Many interesting techniques have been proposed in the literature, which could improve the various functional elements of a space propulsion system, but no radically new approach appeared until 1994. In that year Miguel Alcubierre described a theoretical approach to what appears to be a form of propellantless propulsion capable of reaching the highest speeds (Alcubierre, 1994).

Alcubierre's analysis did neither addressed the problem of the energy required for his propulsion system, nor he explained the precise nature of the propulsor itself.

The estimations of the energy required for a warp drive have changed from an amount comparable to ten times the total energy content of the universe to an amount of few solar masses.

The hardware of the warp drive propulsor is still a mystery. For a warp drive, negative energy densities are required and the associated exotic matter is forbidden classically. Negative energy densities may exist in quantum field theories, nevertheless it is not know if these favourable conditions can be created in a non transient form and, with more emphasis, it is not known how they could be created in the space surrounding the vehicle.

Fortunately a simpler approach to propellantless propulsion exists and it can be derived from general relativity like Alcubierre's warp drive.

This paper describes how spacetime could be manipulated with gravitational radiation and how gravitational radiation could be generated. The important issues of the amount of energy required and the detailed structure of the propulsion system are still an open question, nevertheless these issues appear within our research abilities.

Our approach is based on some aspects of general relativity, specifically the existence of gravitational radiation, the properties of colliding beams of gravitational radiation and the possible mechanisms for the generation of gravitational radiation.

## ON THE EXISTENCE OF GRAVITATIONAL RADIATION

At the beginning of the XX century Albert Einstein developed the concepts of relativity and newtonian gravity into a more complete and credible theory of gravitation, general relativity. In general relativity there are 10 quantities that can create gravity: the energy density, three components of momentum density, and six components of stress. There are 10 unknowns, represented by the components of the metric tensor.

The field equations can be written in terms of a set of 10 fields that are components of a symmetric 4x4 matrix $h_{ab}$ representing the deviation of the metric tensor from that of special relativity, the Minkowsky metric of flat spacetime.

$$\left[\nabla^2 - \frac{1}{c^2}\frac{\partial^2}{\partial t^2}\right]h_{ab} = \frac{G}{c^4}(source) \tag{1}$$

The source represents the set of energy densities and stresses that can create the field. This expression substantially describes the gravitational field as distortion of the geometry of spacetime as a function of energy, momentum densities and stresses in a source. If velocities in the source are much smaller than $c$, and $h$ is small compared to 1 (non linear terms in the source negligible) then the Einstein equations reduce to the Newton's equation in and near the source.

General relativity admits solutions of the field equations in the form of waves. Einstein himself calculated the emission of gravitational waves from various sources under a number of restrictive conditions. We have indeed numerous exact solutions obtained using the linearized equations. These solutions are employed for gravitational wave research from astrophysical sources in order to study the emission, propagation and the detection of gravitational waves. They have been also employed for the study of the emission of gravitational radiation from experimental devices.

If the above mentioned linearization is not applicable we are dealing with a problem that could be solved with an ad hoc approximation or by very complex numerical methods. For instance the full scale highly relativistic simulation of inspiralling and merging black holes might require 10 years of supercomputer operations.

Usually a problem can be divided into a number of partial problems to which different techniques apply. For instance the emission of gravitational radiation by an astrophysical source may require ad hoc methods with a nonlinearity expansion, but wave propagation and detection can be studied with the linearized approximation (Thorne, 1980).

Gravitational wave are transverse wave like electromagnetic waves, they differ from e-m waves also because of their quadrupolar nature. With a set of free test particles, a passing gravitational wave will produce a small relative acceleration of the particles and of their local inertial frames. The relative acceleration is described by quadrupole-shaped lines of force. The two possible polarizations are "+" and "×" separated by a π/4 angle.

If a set of test particles is distributed along a circle, with the plane containing the particles orthogonal to the direction of propagation of the wave, the passing wave will change the shape of the circle to that of an ellipse, then to a circle, then to an ellipse rotated by π/2 respect to the previous one, etc. The transverse plane gravitational wave are area preserving and the amplitude of the deformation of the circle of test particles is $h$.

In 1918 Albert Einstein derived the expression for the gravitational wave field as a function of the second time derivative of the quadrupole moment of the source.

$$h_{jk}^{\text{TT}}(t,x) = (2/r)(G/c^4)[\ddot{q}_{jk}(t-r/c)]^{\text{TT}} \tag{2}$$

Where *x* is the location of the observer in a coordinate system centered on the source, *r* is the distance between source and observer and $q_{jk}(t') = \int r(x',t') \cdot (x'_j x'_k - \frac{1}{3} d_{jk} r'^2) d^3 x'$ is the source mass quadrupole moment, with *r* the mass density.

The dimensionless amplitude *h* of the gravitational waves of astrophysical origin that could be detected on earth and with a frequency of about 1 kHz is between $10^{-17}$ and $10^{-22}$. Gravitational waves can be detected by measuring the effects of spacetime distortion on a beam of laser light (interferometric sensor) or the resulting stress on a mechanical resonator (Weber bar detector) (Thorne, 1980). The predicted amplitude for astrophysical sources is so small that researchers are not certain that a direct gravitational wave detection has ever been made. The detection of artificially generated gravitational waves is an even worse challenge, because only ultradense materials rotating at near relativistic regimes could efficiently emit gravitational radiation detectable at great distances. In spite of the difficulties a near field test has been succesfully made (Astone, 1991) using a 8.75 kg rotor with a quadrupole moment of $5.5 \cdot 10^{-3}$ kg m$^2$ rotating at 30000 rpm and located at a distance of 3.5m from the center of the Weber bar gravitational wave antenna Explorer at CERN.

As gravitational radiation transfers energy and momentum, its existence could be inferred by the back-reaction on the source. It has been determined that the back-reaction on a possible binary pulsar could change the orbital period of the system in a detectable and very peculiar way. The study of such a system could also provide a validity check for general relativity. Fortunately in 1973 Taylor and Hulse discovered the first binary pulsar, they discovered the signature of the emission of gravitational radiation in the emitted radio pulses, and verified the validity of general relativity. This work has been recognized with the Nobel prize in Physics in 1993 (Hulse, 1994), (Taylor, 1994).

## ON THE PROPERTIES OF GRAVITATIONAL RADIATION

Two properties of gravitational radiation are of interest for us. The first property is that gravitational radiation can be directly employed for propulsion. The second property is a consequence of the non-linearity of Einstein equations. If the amplitude of the gravitational wave is sufficiently high, this non-linearity is the source of harmonics and coulomb-like components. Moreover, for colliding beams, spacetime may act as a mixer, again with the appearance of coulomb-like components. By comparison, the propagation of an electromagnetic wave in a non linear medium, for instance the ionosphere, may cause charge separation which is the origin of ionization and electric discharges, but it is not useful for propulsion, this is because electromagnetism admits positive and negative charges, but macroscopic objects are usually neutral.

We do not discuss the amplitude of the gravitational wave required for the generation of evident non-linearity effects, we simply observe that these phenomena are compatible with a highly relativistic system, for instance a binary system of neutron stars.

About the possibility of accelerating an object only by its internal motion a very interesting paper has recently been published (Bonnor, 1997). Here the motion of a rocket accelerated by an anisotropic emission of gravitational waves has been studied using approximation methods. These methods do not assume conservation of momentum or ad hoc formulae, the equation of motion are obtained by directly solving the field equations.

The energy loss of the rocket by the emission of gravitational waves has been found to be in agreement with the quadrupole formula. The power loss is:

$$P = \frac{\dddot{Q}_{xx}^2 + \dddot{Q}_{yy}^2 + \dddot{Q}_{zz}^2}{45(c^5/G)}, \tag{3}$$

where $Q_{ii}$ are the quadrupole moment of the source. For a rod with mass M and length L rotating ω times a second the power *P* emitted with gravitational radiation is:

$$P = \frac{2}{45} \frac{M^2 L^4 w^6}{(c^5/G)}, \tag{4}$$

With the rocket at rest for t=0, the velocity V acquired at t=t1 has been found to be:

$$V = \frac{mGa^5}{315c^7}\int_0^{t1} \dot{p}\dot{q}\,\mathrm{d}t,\qquad(5)$$

where *m* is the initial mass of the rocket, *a* some convenient length associated with it and *p*, *q* are function of *t*, the quadrupole moment is here $Q(t)=ma^2h(t)$, and the octupole moment is $O(t)=ma^3k(t)$, moreover $p(x)=\mathrm{d}^2h(x)/\mathrm{d}x^2$ and $q(x)=\mathrm{d}^3k(x)/\mathrm{d}x^3$. Both expressions contain terms indicating that, with today's knowledge, only astrophysical objects could emit gravitational radiation capable of producing a detectable effect, but we must take into account that in general relativity the definition of the energy transported by a plane gravitational wave is an open question.

About the second property of gravitational radiation which could be of interest for space travel, the results obtained after 30 years of research may be briefly described as follows. The interaction of two impulsive plane waves with infinite wavefronts starts a self focusing process which is believed to end with the creation of a spacetime singularity regardless of the amplitude of the wave (Szekeres, 1972). The time required for the creation of the singularity is a function of the amplitude of the waves *A* and the relative polarization ***a*** of the two waves (Ferrari, 1988).

$$\Delta t = \frac{1}{A^2}\sqrt{1+\sin\boldsymbol{a}}\qquad(6)$$

These results can also be applied to the more realistic case of beam-like gravitational waves (Ferrari, Pendenza, Veneziano, 1988), and are confirmed by a work describing the interaction of two graviton beams (Veneziano, 1987). The collision of gravitational plane wave may produce a curvature singularity or a coordinate singularity, where the radiation is completely converted in a coulomb-like gravitational field. We may now make the conjecture that a single perfectly focused beam of gravitational radiation might produce a spacetime singularity at the focus regardless the amplitude of the wave, moreover in the eventuality of optical imperfections of the focused beam, they could be spontaneously reduced by the behaviour of the collision process, this property should improve with the amplitude and the high frequency content of the wave.

The solutions of the collision problem correspond to a class of black hole solutions, the similarity of the solutions does not imply that the two physical system are identical or identically stable when the external conditions are changed, but we might expect similar effects. Although the similarity could be simply due the precise mathematical description of the collision problem, the possibility of creating a mini black hole cannot be excluded with today knowledge, and this is the main safety concern if attempts are made for creating these conditions in a laboratory.

Again, theories give curious results, in general relativity non-linearity is associated to highly relativistic systems like extreme astrophysical objects, but the interaction of gravitational waves seems capable of reaching these extreme conditions regardless the amplitude of the wave.

We have seen that the mutual interaction of gravitational waves would cause the appearance of a rectified wave, accompanied by a coulomb-like gravitational field. If this field is created outside a spacecraft, the spacecraft would free-fall towards the distorsion. Using the classical picture for the spacetime distortion created by a mass, our spacecraft would follow a depression in spacetime. The moving depression would in turn emit energy as gravitational waves like a moving mass.

Using the famous Einstein equation we observe that energy density (matter) is not an efficient source of gravity because of $m=E/c^2$, and for two equal massive particles the internal rest energy is much higher than the integral gravitational energy, moreover moving this source of gravity to a different location requires the transportation of this energy density. Instead, gravitational waves can be directly and completely converted into a gravitational field which should follow the focus point of the beam(s). The source(s) of the beam(s) could be onboard the spacecraft.

# LABORATORY GENERATION OF GRAVITATIONAL RADIATION

Historically, the emission of gravitational radiation has been studied in astrophysical systems, as reported in the second section of this paper, and a quite large literature exists on the subject. Instead, the laboratory generation of gravitational radiation is still in its theoretical stage of development except, perhaps, the rotor employed for the calibration of a GW antenna (Astone, 1991), which has indeed produced experimental results for a near field detection.

Because we expect that gravitational radiation should be generated starting from a known energy source, we observe that an important parameter certainly is conversion efficiency i.e. the ratio between the output power in gravitational radiation and the input power. The difference between the input and output power has to be dissipated and this could be a serious collateral problem.

Moreover, from eq. 4 we observe that among the many parameters, the amplitude is influenced by $\omega^6$, leading to the idea that a higher frequency could improve the output of the generator.

Combining the request of higher frequencies and low losses leads to the idea that a microscopic source of gravitational radiation could be the preferred source for our application.

The possible structure, the computed output power and conversion efficiency of classical and quantum sources of gravitational radiation are shortly described in the following sub-sections.

## Gravitational Radiation from Classical Systems

Three non-quantum mechanism can be identified for the possible laboratory generation of high frequency gravitational radiation (Pinto, 1988): the coherent EM-GW converter, the EM-pulsed source and the photon/phonon pumped generator. The possibility of arranging several elementary generators in an array will give a beamed emission.

In EM-GW converter the stress-energy tensor of the EM field is the source of the gravitational radiation.

The EM-GW converter has been studied on the basis of a cylindrical EM resonator with a static axial bias magnetic field $H_0$.

Using the $TE_{111}$ resonant mode the power dissipated by the walls and the power emitted by gravitational waves has been computed (Pinto, 1988).

$$P_{walls} = \frac{\omega_{111} W_{em}}{Q_{111}} \tag{7}$$

$$P_{gw} = \frac{G(H_O H_{111})^2 \sin\theta_c \Delta\theta}{(2\pi R/p'_{11})^2 (1+2.912(R/d)^2)^{-1}} \tag{8}$$

In which it has been assumed the gw radiation contained in the conical beam $\theta_c - \Delta\theta/2 < \theta < \theta_c + \Delta\theta/2$.

Using $H_0 = H_{111} = 10^5$ Gauss, $l = 1m$, $d/R = 10$, we obtain $P_{gw} \approx 10^{-17}$ W, with $P_{input} \approx P_{walls} \approx 10^9$ W.

The EM-pulsed source consists in a short solenoid or permanent magnet and a TEM transmission line which traverse the magnetic field generated by the solenoid. Pulses of EM energy are then sent along the transmission line. The amplitude $h \approx 10^{-34}$ is expected from this generator if employing ring transmission lines with a radius of 10 km and EM pulse generators with power of several MW.

The photon/phonon pumped GW generator consists of an array of piezoelectric plates excited by a UHF-SHF modulated laser beam.

An expression of the GW output power of this generator has been obtained (Pinto, 1988):

$$P_{gw} \approx 4 \cdot 10^{-21} \left[\frac{v_s}{c}\right]^3 \left[\frac{1(cm)}{l}\right] \left[\frac{S_w}{10(cm^2)}\right] \left[\frac{\rho_0}{5(g/cm^3)}\right] \left[\frac{Q_{ac}}{100}\right] \left[\frac{P_{ac}}{10^4(W)}\right] W \tag{9}$$

Where $Q_{ac}$ is the quality factor of the plates, $P_{ac}$ the input power, $S_w$ and $l$ respectively area and length of the plates, $v_s$ and $c$ the speed of sound and the speed of light respectively.

The computed conversion efficiency for typical $v_s/c$ ratio is about four order of magnitude lower than that of a pure EM converter. If $v_s \gg c$ than the efficiency of this converter could become about 13 orders of magnitude higher than the pure EM converter.

By comparison to the EM-GW converter we have, for coherent sources:

$$\frac{h_{mech}}{h_{em}} = \frac{r_{mech}}{r_{em}} \left[ \frac{v_{sound}}{c} \right]^5 \tag{10}$$

For completeness, we have that typical $v_{sound}/c \gg 10^{-5 \sim -6}$, and we see that EM fields of about $10^8$ V/cm produce a mass density equivalent of that of water.

## Gravitational Radiation from Quantum Systems

Like electromagnetic radiation, gravitational radiation could be emitted by quantum transitions. We introduce this concept with the simple analogy to the well known binary pulsar GW source. By this analogy the ideal laboratory source of gravitational radiation could be a couple of almost identical orbiting objects with nuclear matter density and with electric charge, which will give us the ability of controlling them with an electromagnetic field. Being the scale factor in principle not relevant for the efficiency of the EM-GW converter, and looking for a high frequency array of such objects, we could reduce the scale factor reaching, for instance, the size of Cooper's pairs, which certainly satisfy our initial requirements.

At atomic scale the emission of a quantum of gravitational radiation, the graviton, is accompanied by a $L=2$ transition in the quantized angular momentum of the emitting system.

In (Halpern, 1964) the investigation of the interaction of the gravitational fields with microscopic systems has been extended to the nuclear and molecular phenomena, with the interesting result that the gravitational interactions have here a greater significance than at a macroscopic level, where gravitational radiation is extremely difficult to generate and detect. The multipole expansion of the gravitational radiation field resulting from periodically oscillating sources has been performed in full analogy with the method used for the electromagnetic fields, thus formally reproducing a successful and experimentally tested methodology. According to (Halpern, 1964) and (Halpern, 1968) atomic transitions for which the orbital quantum number L changes by + or – 2, and for which the total quantum number J changes by 0 or + or – 2 are gravitational quadrupolar transitions and are permitted for the emission of gravitons, while the emission of photons is forbidden. It has been found that atomic transitions from orbitals 3d to 1s, 3d to 2s and 3d to 3s are possible candidates for transitions, which may be applicable for the generation of gravitational radiation by atoms of a suitable material. The material could be pumped by photons and let decay gravitationally. Unfortunately the gravitational transition probability is much lower than in the electromagnetic case; this ratio for matrix elements of equal structures is of the order of:

$$R = \chi^2 (mc)^2 / e^2 \tag{11}$$

where $\chi^2$ is defined in (Halpern, 1964), $e$ is the charge and m the mass of the emitting particle. For a proton this ratio is $1.6 \cdot 10^{-36}$, but it could be about $10^4$ times larger for molecular transitions. Halpern and Laurent first looked for a natural source of high frequency gravitational radiation, they computed that the energy of gravitons involved in some possible stellar processes was very high, 14.4 keV for $^{57}$Fe in the sun and 16.1 MeV from supernovae (Halpern, 1964). Discussing the possibility of a stimulated emission, they also suggested the physical structure of the gravitational counterpart of the laser, called a "gaser". The device structure was a cavity-less single pass device similar to X-ray lasers of today. Again, the extremely low probability of graviton absorption and emission indicates that gasers might not be possible within simple atomic systems.

We have seen that if we pump a given quantum system to a gravitationally excited state with electromagnetic radiation, the resulting state is also an electromagnetically excited state and the probability of electromagnetic emission is higher than the gravitational one by a factor of about $10^{36}$, leading to the conclusion that a different approach is required.

We now imagine a binary quantum system in which we abruptly change the attraction force between the two equal particles composing the system, the resulting probability of states changes accordingly, and if the appropriate change has taken place, an induced emission of a graviton could result. Measurements of the quantized angular momentum of Cooper's pairs in different superconducting materials have been made, and they are compatible with above mentioned possibility. In fact, recently (Harlingen, 1995), (Kouznetsov, 1997), (Sigrist, 1995), (Ding, 1996), (Barret, 1991) it has been experimentally observed the existence of only two different symmetries of the order parameter in low $T_c$ and high $T_c$ superconductors, a symmetry with a *s*-wave component (LTSC and YBCO HTSC) and a symmetry with a *d*-wave component (HTSC) (Kouznetsov, 1997), therefore we know that Cooper-pairs are in *s*-orbitals and *d*-orbitals respectively. We can also predict that when Cooper-pairs move under non equilibrium conditions, i.e. under the effect of magnetic fields, from a superconductor where the symmetry of the order parameter is of type *d* to a superconductor where this symmetry is of type *s*, they are subject to transition and loose energy by the emission of a particle with a spin of 2. A related phenomenon, which is important for the estimation of the binding energy, is the observed emission of wide band THz electromagnetic radiation in a YBCO/insulator/normal-metal junction. This emission is not related to the Josephson effect, and it has been found to be originated by the recombination of Cooper-pairs and quasiparticles at the interface with the non-superconducting material that gives a channel to photon and phonon recombination (Lee, 1998); although the power of the emission was very small, the experiment shows that the electron binding energy is actually released at the interface and a measure of this energy is given. A comprehensive discussion on the symmetry of pairing states in both conventional and high Tc superconductors can be found in (Harlingen, 1995) and therein references.

A junction between a *s*-wave and a *d*-wave superconductor (SDS junction) can be here defined with the purpose of inserting the junction in an electrical circuit and studying the maximum possible emission of radiation. We could estimate the maximum amplitude of the radiation emitted by a SDS junction employing an energy balance equation. Keeping the SDS junction at a temperature much below Tcs and Tcd, the binding energy that is released by a single transition could be a fraction of: $|T_{cd} - T_{cs}|k_B$, where $T_{cd}$ is the critical temperature of the *d*-wave superconductor, $T_{cs}$ is the critical temperature of the *s*-wave superconductor and $k_B$ the Boltzmann constant, considering that the electron binding energy is proportional to the critical temperature.

If we make the hypothesis that this fraction is a factor of one, we may write the usual energy balance equation: $|T_{cd} - T_{cs}|k_B = h\nu$, obtaining frequencies of the order of hundreds of GHz, which are near the frequencies observed in (Lee, 1998). The maximum power emitted by the process could be found with the hypothesis of currents about the critical currents of most superconductors, and of about 10 kA/cm$^2$, this current density may also suppress the Josephson effect on the Bose condensate. Introducing the charge of the electron we obtain a power density of:

$$|T_{cd} - T_{cs}|k_B \left(10^4 / 3.2 \cdot 10^{-19}\right) \quad ; \tag{12}$$

which is of the order of ten W/cm$^2$.

Coherent and collimated emission orthogonal to the plane of the junction is expected because transitions are between two Cooper pairs reservoir where pairs are all condensed in two quantum states so that the frequency and phase of the radiation emitted by each transition is rigorously the same for each event.

A detailed analysis of SDS junctions including possible gravitational effects has been not yet performed, mainly because no satisfactory theory exists to explain electron pairing in high $T_c$ superconductors. Therefore whether these induced *L*=2 transitions are associated to gravitons remains to be investigated both theoretically and experimentally.

Despite the lack of strong research efforts towards the here presented possibility, some kind of SDS junction have been recently studied in order to gain some further knowledge in high temperature superconductivity (Moessle).

M.Moessle and R. Kleiner have experimentally found that a nonzero supercurrent in a c-axis Pb/BSCCO Josephson tunnel junction can be observed and cannot be explained by trivial reasons. It appeared that, an s-wave component of the superconducting order parameter has to be present also in BSCCO, which is a pure d-wave superconductor. Moessle and Kleiner reported that in a strict sense, the data show the existence of the s component only in the vicinity of the interface and at temperatures below the Tc of Pb.

A theory on SDS junction, which does not include gravitational effects has been developed by Yukio Tanaka His paper presents a theory which predicts a supercurrent in s/i/d junctions, using the previous theories by Arnold, Furusaki and Tsukada which include infinite orders of the tunneling process. Josephson current between d wave/insulator/s wave planar-contact junction (d/s junction) is calculated. In such cases, critical current Ic is proportional to Tc-T near the transition temperature Tc. The basic argument is that s wavefunction and d wavefunction are orthogonal and no supercurrent can be predicted for a SDS junction unless a particle associated to a kind of radiation is emitted or absorbed. We identify this possible particle with the graviton, which has the appropriate spin.

Halpern and Laurent's theory predicts an extremely low probability of emission of gravitons on a general basis, nevertheless this is a linearized theory, we can try to be more realistic by considering the relativistic mass of the electron in the tunneling process: $m=m_0/\sqrt{(1-v^2/c^2)}$. Here an additional difficulty arises because electrons apparently might traverse tunnel junctions at the speed of light making its mass infinite for and infinitesimal time. The ratio in eq. 11 could become infinite for an infinitesimal time and much higher than one for an interval of time sufficient for the emission of a graviton.

## CONCLUSION

This paper has shortly described the most relevant elements of what the author believes could become a new propulsion technique. The theoretical background is that of general relativity and the main subject of the paper is gravitational radiation from man made sources. With reference to a wide collection of theoretical papers and few experimental one, it has been shown that gravitational radiation can be employed for space propulsion. Moreover it has been shown that gravitational radiation can be generated by artificial means and a new hypothesis on its possible emission from a quantum system has been proposed. Further developments are expected by a deeper analysis of the collision problem and the study of gravitational transitions in some promising materials.

## REFERENCES

This paper contains and extends the paper:
Fontana Giorgio, "Gravitational radiation and its application to space travel", Proceedings of Space Technology and Applications International Forum-2000, STAIF-2000 "Conference on Enabling Technology and Required Scientific Developments for Interstellar Missions", Jan 30 - Feb 3, 2000, Albuquerque, NM. AIP Conference Proceedings 504 1085-1092.

References:

Alcubierre Miguel, "The Warp Drive: Hyper-Fast Travel Within General Relativity", *Class. Quantum Grav*,

**11**, 1994, pp. L73-L77

Astone et…,"Evaluation and Preliminary Measurement of the Interaction of a Dynamical Gravitational Near Field with a Cryogenic G.W. Antenna", *Zeischrift fuer Physik C,* **50**, 1991, pp. 21-29

Barret S.E., Martindale J.A., Durand D.J., Pennington C.P., Slichter C.P., Friedmann T.A., Rice J.P., and Ginsberg D.M., "Anomalous Behavior of Nuclear Spin-Lattice Relaxation Rates in YBa$_2$Cu$_3$O$_7$ below Tc", *Phys. Rev. Lett*. 1991 , **66**, pp**.** 108-111.

Bonnor W. B. and Piper M. S., "The Gravitational Wave Rocket", *Class. Quantum Grav.* ,**14**, 1997, pp. 2895-2904.

Ding H. et al. , "Angle-resolved Photoemission Spectroscopy Study of the Superconducting Gap Anisotropy in Bi$_2$Sr$_2$CaCu$_2$O$_{8+x}$. ", *Physical Review B*., **54**, N. 14, 1 Oct. 1996-II, pp. R9678-R9681

Ferrari V., Pendenza P., and Veneziano G., "Beam-like Gravitational Waves and Their Geodesics", *General Relativity and Gravitation,* **20**, No 11, 1988, pp. 1185-1191

Ferrari Valeria, "Focusing Process in the Collision of Gravitational Plane Waves", *Physical Review D,* **37**, No10, 15 May 1988, pp. 3061-3064

Halpern L., Jouvet B., "On the Stimulated Photon-Graviton Conversion by an Electromagnetic Field*", Annales H. Poincare*, **VIII**, NA1, 1968 pp. 25-42.

Halpern L., Laurent B., "On the Gravitational Radiation of Microscopic Systems", *IL Nuovo Cimento,* **XXXIII**, N. 3, 1964, pp. 728-751

Harlingen D.J., "Phase-sensitive Tests of the Symmetry of the Pairing State in High-temperature Superconductors- Evidence for d Symmetry", *Reviews of Modern Physics*, **67** No. 2, 1995, pp. , 515-535.

Hulse Russel A., "The Discovery of the Binary Pulsar", *Reviews of Modern Physics*, **66**, No. 3, July 1994, pp. 699-710,.

Kouznetsov K.A. et al., "*c*-axis Josephson Tunneling between YBa$_2$Cu$_3$O$_{7-x}$ and Pb: Direct Evidence for Mixed Order Parameter Symmetry in a High Tc Superconductor", *Physical Review Letters*, **79**, 20 Oct. 1997, pp. 3050-3053,.

Lee Kiejin, Iguochi Ienari, Arie Hiroyuki and Kume Eiji, "Nonequilibrium Microwave Emission from DC-Biased High Tc YBa$_2$Cu$_3$O$_{7-y}$ Junctions", *Jpn. J. Appl Phys.* ,**37** Part 2, No. 3A, 1 March (1998) pp. L 278-280.

M.Moessle and R. Kleiner – "*c*-axis Josephson tunneling between BSCCO and Pb", PHYSICAL REVIEW B, VOL. 59 N. 6, 4486-4496. 1 FEB. 1999

Pinto I.M. and .Rotoli G, "Laboratory Generation of Gravitational Waves ?" *Proceedings of the 8$^{th}$ Italian Conference on General Relativity and Gravitational Physics",* Cavalese (Trento) August 30 – September 3, 1988 World Scientific – Singapore p 560-573

Sigrist M. and Rice T.M., "Unusual Paramagnetic Phenomena in Granular High-temperature Superconductors- A Consequence of *d*-wave Pairing ? ", *Reviews of Modern Physics*, **67**, No. 2, 1995, pp. 503-513.

Szekeres P., "Colliding Plane Gravitational Waves", *J. Math. Phys,* **13**, No. 3, March 1972, pp. 286-294

Tanaka Yukio, "Josephson Effect between s Wave and d Wave Superconductors", PHYSICAL REVIEW LETTERS, VOL. 72 N. 24, 3871-3874. 13 JUN. 1994.

Taylor Joseph H., Jr, "Binary pulsar and relativistic gravity", *Reviews of Modern Physics*, **66**, No. 3, July 1994, pp. 711-719.

Thorne Kip S., "Gravitational-wave Research: Current Status and Future Prospects", *Reviews of Modern Physics*, **52**, No. 2, Part I, April 1980, pp. 285-297.

Thorne Kip S., "Multipole Expansion of Gravitational Radiation", *Reviews of Modern Physics*, **52**, No. 2, Part I, April 1980, pp. 299-339

Veneziano G., "Mutual Focusing of Graviton Beams", *Modern Physics Letters A,* **2**, No 11, 1987, pp. 899-903